\documentclass[10pt]{article} 
\usepackage[accepted]{rlc}

\usepackage{amssymb}            
\usepackage{mathtools}          
\usepackage{mathrsfs}           
\mathtoolsset{showonlyrefs}     
\usepackage{graphicx}           
\usepackage{subcaption}         
\usepackage[space]{grffile}     
\usepackage{url}                

\usepackage{amsthm}
\newtheorem{proposition}{Proposition}

\newtheorem{theorem}{Theorem}
\newtheorem{assumption}{Assumption}

\usepackage{xcolor}         

\title{Agent-Temporal Credit Assignment for Optimal Policy Preservation in Sparse Multi-Agent Reinforcement Learning}

\author{Aditya Kapoor  \\
    Tata Consultancy Services, Mumbai \\
    \And
    Sushant Swamy \\
    Birla Institute of Technology and Science, Goa \\
    \And
    Kale-ab Tessera \\
    University of Edinburgh \\
    \And
    Mayank Baranwal \\
    Tata Consultancy Services, Mumbai \\
    \And
    Mingfei Sun \\
    University of Manchester \\
    \And
    Harshad Khadilkar \\
    Indian Institute of Technology, Bombay \\
    \And
    Stefano V. Albrecht \\
    University of Edinburgh \\
    }

\begin{document}

\maketitle

\begin{abstract}

In multi-agent environments, agents often struggle to learn optimal policies due to sparse or delayed global rewards, particularly in long-horizon tasks where it is challenging to evaluate actions at intermediate time steps. We introduce Temporal-Agent Reward Redistribution (TAR$^2$), a novel approach designed to address the agent-temporal credit assignment problem by redistributing sparse rewards both temporally and across agents. TAR$^2$ decomposes sparse global rewards into time-step-specific rewards and calculates agent-specific contributions to these rewards. We theoretically prove that TAR$^2$ is equivalent to potential-based reward shaping, ensuring that the optimal policy remains unchanged. Empirical results demonstrate that TAR$^2$ stabilizes and accelerates the learning process. Additionally, we show that when TAR$^2$ is integrated with single-agent reinforcement learning algorithms, it performs as well as or better than traditional multi-agent reinforcement learning methods.
\end{abstract}

\section{Introduction}
\label{sec:introduction}

In cooperative multi-agent reinforcement learning (MARL), multiple autonomous agents learn to interact and collaborate to execute tasks in a shared environment by maximizing a global return \citep{busoniu_2008}. MARL has shown considerable potential in solving decentralized partially observable Markov decision processes (Dec-POMDPs) \citep{Oliehoek2016ACI, amato2024partial, zhang2021multi}, where each agent has access to only local information (partial observation) and need to select actions based on their local action-observation (or sometimes only observation) histories to maximize the global (team) return. Applications of MARL include complex video games such as StarCraft-II \citep{Vinyals2019GrandmasterLI}, Defense of the Ancients (DOTA) \citep{berner2019dota}, Google Football \citep{kurach2020google}, and Capture the Flag (CTF) \citep{jaderberg2019human}, and real world applications of warehouse logistics \citep{krnjaic2022scalable, RTAW}, e-commerce \citep{shelke2023multi, baer2019}, robotics \citep{sartoretti2019primal, damani2021primal}, and routing problems \citep{zhang2018fully, vinitsky2020optimizing, zhang2023learning}. These applications illustrate the potential of MARL to develop sophisticated strategies and behaviors through coordinated teamwork and collaboration.


Despite these successes, cooperative multi-agent systems face a significant challenge in credit assignment, which is crucial for learning effective policies \cite{foerster2018counterfactual}. Credit assignment in multi-agent systems encompasses two main aspects: \emph{temporal credit assignment} and \emph{agent credit assignment}. Temporal credit assignment involves decomposing sparse, delayed rewards into intermediate time steps within a multi-agent trajectory. The assignment of agent credit focuses on discerning the contribution of each agent to these decomposed temporal rewards \citep{marl-book}. Addressing both aspects is essential for effective learning in cooperative multi-agent systems.


Significant progress has been made in addressing the credit assignment problem with methods such as VDN\citep{sunehag2017value}, QMIX\citep{rashid2020monotonic}, QTRAN\citep{son2019qtran}, COMA\citep{foerster2018counterfactual}, and PRD \citep{freed2021learning}. However, these methods are primarily designed to deal with agent credit assignment and may not be suitable for environments with sparse or delayed rewards \citep{Papoudakis2020BenchmarkingMD, de2020independent}. Additionally, the representations required for effective credit assignment might not align with those needed for learning Q-values or critics. Recent advances in temporal credit assignment have introduced dense reward functions in single-agent settings \citep{arjona2019rudder, ren2021learning, liu2019sequence, gangwani2020learning} and multi-agent settings \citep{xiao2022agent, she2022agent}. These methods aim to learn a Markovian proxy reward function that replaces the environment's sparse rewards with dense rewards. Motivated by this progress, we aim to address the combined challenge of agent and temporal credit assignment in multi-agent tasks with sparse or delayed rewards.

In this paper, we propose Temporal-Agent Reward Redistribution (TAR$^2$), a novel approach to address the problem of agent-temporal credit assignment by learning a reward redistribution function that decomposes sparse environment rewards to each time step of the multi-agent trajectory and further redistributes the temporally decomposed rewards to each agent based on their contributions. We theoretically prove that there exists a class of reward redistribution functions that can be formulated as potential-based reward shaping \citep{ng1999policy, Devlin2011TheoreticalCO}, under which the optimal policies are preserved in the original reward function of the environment. TAR$^2$ extends AREL's \citep{xiao2022agent} reward model that uses a temporal attention module to analyze the influence of state-action tuples along trajectories, followed by an agent attention module to identify the relevance of other agents for each agent. This alternation between the two attention modules allows the reward function to identify agent-specific state-action tuples that are key to the sparse environment rewards received by the multi-agent system. Thus, TAR$^2$ learns agent-specific temporal rewards and enables the use of single agent reinforcement learning (RL) algorithms such as IQL \citep{Tan1997MultiAgentRL}, IAC \citep{foerster2018counterfactual}, and IPPO \citep{schulman2017proximal, de2020independent} to solve multi-agent tasks. This approach separates the problem of credit assignment from learning Q-functions and critics, leveraging the simplicity and scalability of single-agent RL algorithms in complex environments. We empirically demonstrate the sample efficiency of our approach against competitive state-of-the-art baselines on SMACLite \citep{michalski2023smaclite}. 

In summary, our contribution is three folds:-
\begin{itemize}
    \item We introduce TAR$^2$, a reward redistribution model that can be used to assign temporal and agent credit assignment to densify the reward signal.
    \item We theoretically show that TAR$^2$ is equivalent to potential-based reward shaping which ensures that the optimal policy learned using TAR$^2$ is also optimal under the environment's original reward function.
    \item We theoretically prove that any intermediate policy gradient updates under TAR$^2$ and the environment's original reward function share the same direction, ensuring that the trajectory of policy updates remains consistent across both reward functions.
    \item We empirically validate our method on  Alice \& Bob, various battle environments in SMACLite \citep{michalski2023smaclite} and various environment configurations of Google football \citep{kurach2020google}.
\end{itemize}

\section{Related Works}
\label{sec:related-works}

In this section, we review various techniques proposed to address temporal and agent credit assignment in both single-agent and multi-agent systems. We begin by discussing potential-based reward shaping, a method that provides theoretical guarantees of sample-efficient learning of optimal policies in single-agent \citep{ng1999policy} and multi-agent \citep{lu2011policy, Devlin2011TheoreticalCO} settings. The use of potential based shaping methods have shown to accelerate the learning process.

\subsection{Temporal Credit Assignment}
\label{subsec:temporal-credit-assignment}
Temporal credit assignment focuses on decomposing sparse or episodic environment rewards into dense reward functions by attributing credit to each time step in an episode.

RUDDER \citep{arjona2019rudder} and its variants \citep{patil2020align} use contribution analysis to break down episodic rewards into per-time-step rewards by computing the difference between predicted returns at successive time steps. Similarly, \citet{zhang2023grd} perform return-equivalent contribution analysis. \citet{liu2019sequence} leverage auto-regressive architectures from natural language processing, such as Transformers \citep{vaswani2017attention}, to attribute credit to every state-action tuple in the trajectory. Methods by \citet{efroni2021reinforcement} and \citet{ren2021learning} learn proxy reward functions via trajectory smoothing based on reinforcement learning algorithms that utilize least-squares error. \citet{harutyunyan2019hindsight} introduce a new family of algorithms that use new information to assign credit in hindsight. \citet{han2022off} redesign the value function to predict returns for both historical and current steps by approximating these decompositions. \citet{zhu2023towards} propose a bi-level optimization framework to learn a reward redistribution for effective policy learning. These methods have been primarily developed for single-agent settings and may not scale well to multi-agent reinforcement learning (MARL) due to the exponential growth in the joint observation-action space.

In multi-agent settings, recent works have also addressed temporal credit assignment. IRCR \citep{gangwani2020learning} developed a count-based method to learn a proxy reward function for both single and multi-agent settings. AREL \citep{xiao2022agent} uses attention networks to perform reward redistribution, while \citet{she2022agent} employ an attention encoder network followed by a decoder to address both agent and temporal credit assignment in delayed reward settings.

\subsection{Agent Credit Assignment}
\label{subsec:agent-credit-assignment}
Most prior works focus on agent credit assignment in multi-agent systems. \citet{devlin2014potential} and \citet{foerster2018counterfactual} employ difference rewards to assess each agent's contribution to the global reward. Value Decomposition Networks (VDN) \citep{sunehag2017value} decompose the joint value function into agent-specific value functions, assuming additivity. \citet{rashid2020monotonic} introduce monotonicity constraints on the joint Q function to learn individual Q values for each agent. \citet{son2019qtran} generalize this approach to decompose joint Q functions into agent-specific Q functions. \citet{wang2020shapley} leverage Shapley values to model the joint Q function for agent credit assignment. \citet{zhou2020learning} propose an entropy-regularized actor-critic method to efficiently explore multi-agent credit assignment. \citet{freed2021learning} use Transformer attention mechanisms in the critic of an actor-critic method to identify relevant agent subgroups for effective multi-agent credit assignment. However, these techniques do not address temporal credit assignment and are therefore inadequate for learning optimal policies in episodic or highly delayed reward settings.

In summary, while significant progress has been made in addressing either agent or temporal credit assignment, the combined challenge of both remains underexplored. Our proposed Temporal-Agent Reward Redistribution (TAR$^2$) aims to fill this gap by effectively handling both agent and temporal credit assignment, enabling efficient learning in multi-agent environments with sparse or delayed rewards.

\section{Background}
\label{sec:background}


In this section, we introduce our problem setup within the framework of a decentralized partially observable Markov decision process (Dec-POMDP) \citep{Oliehoek2016ACI, amato2024partial}. We describe how agents operate under partial observability and must make decisions based on local observations and histories. We then discuss the episodic multi-agent reinforcement learning (MARL) setting, where agents receive rewards only at the end of each episode, posing a significant challenge for credit assignment. To address this, we explore potential-based reward shaping for multi-agent systems \citep{ng1999policy, Devlin2011TheoreticalCO}, a technique that reshapes the reward function to facilitate learning while preserving the optimal policy. Finally, we analyze how imperfect credit assignment impacts the variance of the policy gradient in multi-agent systems. We show that improper credit distribution among agents leads to high variance in advantage estimates, which in turn exacerbates the learning process and hinders the convergence to optimal policies.

\subsection{Decentralized Partially Observable Markov Decision Processes (Dec-POMDP)}
\label{subsec:decpomdp_def}
A Dec-POMDP is represented by a tuple $\mathcal{M} = (\mathcal{S, A, P, T, O, N}, \mathcal{R}_{\zeta}, \rho_0, \gamma)$ where $s \in \mathcal{S}$ is the environment state space, $a \in \mathcal{A}$ is the joint action space denoted by $\mathcal{A} := \mathcal{A}_1 \times \mathcal{A}_2\times ... \times\mathcal{A}_N$ and $\mathcal{P}: \mathcal{S} \times \mathcal{A} \times \mathcal{S} \to [0,1]$ is the state transition function. $r_{\text{global}, t} \sim \mathcal{R}_{\zeta}( | s_t, a_t): \mathcal{S} \times \mathcal{A} \to \mathbb{R}$ is the global reward shared among agents at every timestep of the trajectory. 
$\rho_0$ is the initial state distribution and $\gamma \in [0, 1]$ is the discount factor. $\pi = \prod_{i=1}^N \pi_i$ is the joint policy of the multi-agent system which comprises of independent agent policies $\pi_i$. 
Each agent $i \in \{1 \dots N\}$ receives an observation $o_i \in \mathcal{O}_i$ from the observation function $\mathcal{T}(s, i): \mathcal{S} \times \mathcal{N} \to O$. Because the state is not directly observable, it is typically beneficial for each agent to remember a history of its observations or observations-actions. $h_{t} \in \mathcal{H}:= \mathcal{H}_1 \dots \mathcal{H}_N$ is the set of agent observation (-action) histories up to the current time step $t$ where $h_{i, t} \in \mathcal{H}_i$ and defined as $h_{i, t} = \{o_{i,1}, a_{i,1}, . . . , o_{i,t}\}$ denotes agent $i$'s history and $h_{\neg i, t}$ is the history of all other agents except agent $i$. At each time step every agent selects an action $a_i \in \mathcal{A}_i$ according to it's policy $\pi_i : H_i \times \mathcal{A}_i \to [0, 1]$. $\tau = \{o_{0, 1}, a_{0, 1}, ... o_{0, N}, a_{0, N} ....... o_{|\tau|, 1}, a_{|\tau|, 1} ... o_{|\tau|, N}, a_{|\tau|, N}\}$ is the multi-agent trajectory where |$\tau$| is the horizon length of the trajectory. 
The goal of the agents is to determine their individual optimal policies that achieve maximum global return
$E_{s_0 \sim \rho_0, s \sim \mathcal{P}, a_i \sim \pi_i ..}\left[\sum_{t=1}^{|\tau|} \gamma^t r_t \right]$.

\subsection{Return Decomposition in Episodic Multi-Agent Reinforcement Learning}
\label{subsec:return_decomposition_episodic_marl}

In most MARL systems, each agent receives a global reward $r_{\text{global}, t}$ after executing the joint action $a_t$ in state $s_t$. However,
in episodic MARL setups, agents only receive a global reward signal from the environment at the end of the trajectory, known as the episodic reward or trajectory return $r_{\text{global}, \text{episodic}}$. The objective in such environments is to maximize the trajectory return, $E_{s_0 \sim \rho_0, s \sim \mathcal{P}, a_i \sim \pi_i ..}(r_{\text{global}, \text{episodic}}(\tau))$. Delayed reward settings introduce significant bias and variance \citep{ng1999policy} during the learning process, exacerbating sample inefficiency.

\subsection{Potential-based reward shaping}
\label{subsec:potential_based_reward_shaping}
\citet{ng1999policy} presented a single-agent reward shaping method to address the credit assignment problem by introducing a potential-based shaping reward to the environment. The combination of the shaping reward with the original reward can enhance the learning performance of a reinforcement learning algorithm and accelerate the convergence to the optimal policy. \citet{Devlin2011TheoreticalCO} and \citet{lu2011policy} extended potential-based reward shaping to multi-agent systems.

\begin{theorem}
\label{theorem: potential_based_rew_shaping_theorem}
Given an $n$-player discounted stochastic game $M = (S, A_1, \ldots, A_n, T, \gamma, R_1, \ldots, R_n)$, we define a transformed $n$-player discounted stochastic game $M' = (S, A_1, \ldots, A_n, T, \gamma, R_1 + F_1, \ldots, R_n + F_n)$, where $F_i \in S \times S$ is a shaping reward function for player $i$. We call $F_i$ a potential-based shaping function if $F_i$ has the form:

\[
F_i(s, s') = \gamma \Phi_i(s') - \Phi_i(s),
\]

where $\Phi_i: S \to \mathbb{R}$ is a potential function. Then, the potential-based shaping function $F_i$ is a necessary and sufficient condition to guarantee the Nash equilibrium policy invariance such that:

\begin{itemize}
    \item \textbf{(Sufficiency)} If $F_i$ ($i = 1, \ldots, n$) is a potential-based shaping function, then every Nash equilibrium policy in $M'$ will also be a Nash equilibrium policy in $M$ (and vice versa).
    \item \textbf{(Necessity)} If $F_i$ ($i = 1, \ldots, n$) is not a potential-based shaping function, then there may exist a transition function $T$ and reward function $R$ such that the Nash equilibrium policy in $M'$ will not be the Nash equilibrium policy in $M$.
\end{itemize}
\end{theorem}

\noindent
In summary, potential-based reward shaping ensures that Nash equilibrium policies are preserved, enhancing learning without altering the strategic dynamics. This principle underpins our proposed reward redistribution method, which we will validate in the following sections, demonstrating its effectiveness in multi-agent reinforcement learning.

\subsection{Impact of Faulty Credit Assignment on Policy Gradient Variance}
\label{subsec:faulty_credit_assignment_impact}
To understand the impact of imperfect credit assignment, we analyze the effect of other agents on the policy gradient update of agent $i$. 
Consider an actor-critic gradient estimate for a multi-agent system in a Dec-POMDP setting, computed using a state-action sample from an arbitrary timestep $t$. We make no assumptions about the policy parameters of the agents in the multi-agent system. Ideally, the policy gradient update for agent $i$ should be computed using

\begin{equation}
    \hat{\nabla}_{\theta_i} J(\theta, h) = \nabla_{\theta_i} \log \pi_i(a_i|h_i) \mathbb{E}_{\neg h_i, \neg a_i}\left[A(h, a)\right]
\end{equation}

Computing $\mathbb{E}_{\neg h_i, \neg a_i}\left[A(h, a)\right]$ ischallenging to compute due to the high dimensionality, dependency on other agents, and the computational complexity involved in accurately modeling and estimating the interdependent histories and actions of multiple agents. 
However, in practice multi-agent policy-gradient algorithms like MAPPO \citep{yu2022surprising}, MADDPG \citep{lowe2017multi} etc employ $A(h, a)$ to compute the policy gradient update for each agent. As a result, the credit assignment problem manifests as high variance in advantage estimates, leading to slower learning because of noisier policy gradient estimates. 



Multi-agent policy gradient methods approximate the true \textit{advantage} by computing $\hat{A}$, which is actually a stochastic advantage estimation of taking a joint action $a$ while observing the joint agent history $h$, and following the joint policy $\pi$. The advantage function is typically defined as $A^{\pi}(s,a) = Q^{\pi}(s, a) - V^{\pi}(s)$, where $Q^{\pi}(s,a)$ and $V^{\pi}(s)$ are the state-action value function and state-value function, respectively \citep{sutton1998introduction}. However, in practice the state-action value function and state-value function are approximated using $\hat{Q}^{\pi}(h,a)$ and $\hat{V}^{\pi}(h)$ in Dec-POMPDs and there are many ways to compute \(\hat{A}\), generally all involving some error, as the true value functions are unknown \citep{sutton1998introduction, gae}.. Intuitively, this is the centralized advantage function which measures how much better it is to select a joint action $a$ than a random action from the joint policy $\pi$, while in state $s$. Besides, to update the policy of agent $i$, we need to compute the advantage of selecting action $a_i$, taking into account the specific contribution and context of agent $i$ within the multi-agent system. Perfect credit assignment would be possible if the advantage function could be computed perfectly for each agent, as it directly measures how a particular action of an agent impacted the total reward obtained by the group. 

The conditional variance of $\mathrm{Var}(\hat{\nabla}_{\theta} J|h,a)$, given $h$ and $a$, is proportional to the variance of $\hat{A}$. While this statement typically implies multiple samples are considered to estimate the variance accurately, we can gain insight into the variance introduced by the contributions of other agents by initially focusing on a single sample scenario.
\begin{equation}
    \mathrm{Var}(\hat{\nabla}_{\theta} J|h,a) = \left( \nabla_{\theta} \log \pi(a_i|h_i) \right)\left( \nabla_{\theta} \log \pi(a_i|h_i) \right)^T \mathrm{Var}(\hat{A}|h,a). \\
\end{equation}
It is therefore evident that the variance of the policy gradient update is directly proportional to the variance of the advantage estimate: $\mathrm{Var}(\hat{\nabla}_{\theta} J|h,a) \propto \mathrm{Var}(\hat{A}|h,a)$. \\
Let us analyze the variance of the advantage function in cooperative multi-agent setting,

\begin{align}
    \mathcal{A}(h, a) &= \mathcal{Q}(h, a) - \mathcal{V}(h) \\
    \intertext{Let us assume that agent $i$'s reward contribution at an arbitrary time-step $t$ is denoted by $r_{i,t}$ and the episodic reward described in subsection \ref{subsec:return_decomposition_episodic_marl} can be derived using $r_{\text{global}, \text{episodic}}(\tau) = \sum_{t=1}^{|\tau|}\sum_{i=1}^N r_{i,t} | h, a$. From this we can rewrite $\mathcal{Q}(h, a)$ and $\mathcal{V}(h)$ as $\mathcal{Q}(h, a) = \mathrm{E}_{s_0 \sim \rho_0, s \sim \mathcal{P}, a_i \sim \pi_i ..}[\sum_{t=1}^{|\tau|}\sum_{i=1}^N r_{i,t} | h, a]$ and $\mathcal{V}(h) = \mathrm{E}_{\pi}[\mathcal{Q}(h, a)]$}
    \mathcal{A}(h, a) &= \mathrm{E}_{s_0 \sim \rho_0, s \sim \mathcal{P}, a_i \sim \pi_i ..}[\sum_{t=1}^{|\tau|}\sum_{i=1}^N r_{i,t} | h, a] - \mathrm{E}_{\pi}[\mathrm{E}_{s_0 \sim \rho_0, s \sim \mathcal{P}, a_i \sim \pi_i ..}[\sum_{t=1}^{|\tau|}\sum_{i=1}^N r_{i,t} | h]] \\
    \intertext{Based on the linearity of expectations on $\sum_{t=1}^{|\tau|}\sum_{i=1}^N r_{j,t} = \sum_{t=1}^{|\tau|} r_{i,t} + \sum_{t=1}^{|\tau|}\sum_{j \neq i}^N r_{j,t}$}
    \mathcal{A}(h, a) &= (\mathrm{E}_{s_0 \sim \rho_0, s \sim \mathcal{P}, a_i \sim \pi_i ..}[\sum_{t=1}^{|\tau|} r_{i,t} | h, a] + \mathrm{E}_{s_0 \sim \rho_0, s \sim \mathcal{P}, a_i \sim \pi_i ..}[\sum_{t=1}^{|\tau|}\sum_{j \neq i}^N r_{j,t} | h, a]) \notag\\
    &\quad - (\mathrm{E}_{\pi}[\mathrm{E}_{s_0 \sim \rho_0, s \sim \mathcal{P}, a_i \sim \pi_i ..}[\sum_{t=1}^{|\tau|} r_{i,t} | h]] + \mathrm{E}_{\pi}[\mathrm{E}_{s_0 \sim \rho_0, s \sim \mathcal{P}, a_i \sim \pi_i ..}[\sum_{t=1}^{|\tau|}\sum_{j \neq i}^N r_{j,t} | h]]) \\
    \mathcal{A}(h, a) &= (\mathrm{E}_{s_0 \sim \rho_0, s \sim \mathcal{P}, a_i \sim \pi_i ..}[\sum_{t=1}^{|\tau|} r_{i,t} | h, a] - \mathrm{E}_{\pi}[\mathrm{E}_{s_0 \sim \rho_0, s \sim \mathcal{P}, a_i \sim \pi_i ..}[\sum_{t=1}^{|\tau|} r_{i,t} | h]]) \notag\\ 
    &\quad- (\mathrm{E}_{s_0 \sim \rho_0, s \sim \mathcal{P}, a_i \sim \pi_i ..}[\sum_{t=1}^{|\tau|}\sum_{j \neq i}^N r_{j,t} | h, a] - \mathrm{E}_{\pi}[\mathrm{E}_{s_0 \sim \rho_0, s \sim \mathcal{P}, a_i \sim \pi_i ..}[\sum_{t=1}^{|\tau|}\sum_{j \neq i}^N r_{j,t} | h]])
    \intertext{The advantage estimate considering only the contribution of agent $i$ is $\mathcal{A}_i = \mathrm{E}_{s_0 \sim \rho_0, s \sim \mathcal{P}, a_i \sim \pi_i ..}[\sum_{t=1}^{|\tau|} r_{i,t} | h, a] - \mathrm{E}_{\pi}[\mathrm{E}_{s_0 \sim \rho_0, s \sim \mathcal{P}, a_i \sim \pi_i ..}[\sum_{t=1}^{|\tau|} r_{i,t} | h]]$ the only advantage term that should be considered while calculating the policy gradient update for agent $i$ whereas the advantage estimate due to other agent contributions $\mathcal{A}_{\neg i} = \mathrm{E}_{s_0 \sim \rho_0, s \sim \mathcal{P}, a_i \sim \pi_i ..}[\sum_{t=1}^{|\tau|}\sum_{j \neq i}^N r_{j,t} | h, a] - \mathrm{E}_{\pi}[\mathrm{E}_{s_0 \sim \rho_0, s \sim \mathcal{P}, a_i \sim \pi_i ..}[\sum_{t=1}^{|\tau|}\sum_{j \neq i}^N r_{j,t} | h]]$ act as noise. Thus,}
    \mathcal{A}(h, a) &= \mathcal{A}_{i}(h, a) + \mathcal{A}_{\neg i}(h, a) \\
    \intertext{Using variance of the sum of random variables,}
    \mathrm{Var}(\mathcal{A}(h, a)) &= \mathrm{Var}(\mathcal{A}_{i}(h, a)) + \mathrm{Var}(\mathcal{A}_{\neg i}(h, a)) + 2 \mathrm{Cov}(\mathcal{A}_{i}(h, a), \mathcal{A}_{\neg i}(h, a))
    \intertext{To express the equation in terms of variance, we use the Cauchy-Schwarz inequality, which states that for any two random variables $\mathcal{A}_{i}$ and $\mathcal{A}_{\neg i}$:}
    \mathrm{Cov}(\mathcal{A}_{i}(h, a), \mathcal{A}_{\neg i}(h, a)) &\le \sqrt{\mathrm{Var}(\mathcal{A}_{i}(h, a)) \mathrm{Var}(\mathcal{A}_{\neg i}(h, a))}
    \intertext{By substituting this inequality, we get an upper bound on our equation,}
    \mathrm{Var}(\mathcal{A}(h, a)) &\le \mathrm{Var}(\mathcal{A}_{i}(h, a)) + \mathrm{Var}(\mathcal{A}_{\neg i}(h, a)) + 2 \sqrt{\mathrm{Var}(\mathcal{A}_{i}(h, a)) \mathrm{Var}(\mathcal{A}_{\neg i}(h, a))} \\
    \mathrm{Var}(\mathcal{A}(h, a)) &\le (\sqrt{\mathrm{Var}(\mathcal{A}_{i}(h, a))} + \sqrt{\mathrm{Var}(\mathcal{A}_{\neg i}(h, a))})^2
\end{align}
The above equation shows that the variance of the policy gradient update grows approximately linearly with the number of agents in the multi-agent system. This increase in variance reduces the signal-to-noise ratio of the policy gradient, necessitating more updates for effective learning. Proper credit assignment can mitigate this issue by enhancing the signal-to-noise ratio, thereby facilitating more sample-efficient learning.

\section{Method}

\subsection{Definition of reward redistribution function}
\label{subsec:rrf_def}
In this paper, we address the challenge of temporal and agent credit assignment in fully cooperative multi-agent systems with episodic global rewards. Our goal is to learn a reward redistribution function that preserves the optimal policy of the original reward function of the environment. We aim to achieve this by defining a reward redistribution function that decomposes the episodic trajectory reward, also known as trajectory return, to each agent based on their contribution to the team's outcome at every time step. 

We premise that the joint history and action at the final timestep of the multi-agent trajectory serve as a good proxy for predicting the episodic global reward. Consequently, we predict the per timestep reward for each agent by assessing the contribution of its state-action tuple towards generating this joint history of the multi-agent system.

\begin{assumption}
\label{assumption: rew_red_ass}
The reward redistribution function $r_{i, t}$, which assigns the reward received by agent $i$ at time step $t$, is determined by analyzing the importance of each state-action tuple against the joint history and action at the final timestep that predicts the episodic global reward.
\begin{equation}
r_{\text{global}, \text{episodic}}(\tau) = \sum_{i=1}^N \sum_{t=1}^{|\tau|} r_{i, t}(h_{i, t}, a_{i, t}, h_{\neg i, t}, a_{\neg i, t}, h_{|\tau|}, a_{|\tau}),
\label{eq:rew_red_assumption}
\end{equation}
\end{assumption}

This assumption aligns with the framework adopted by prior works \citep{xiao2022agent, ren2021learning, efroni2021reinforcement} which also assumes that the episodic return has some structure in nature, e.g., a sum-decomposable form.

\subsection{Assembling the reward function}
\label{subsec:assembling_reward_func}
We propose a method to redistribute the trajectory returns temporally, assigning credit to each time step in the multi-agent trajectory. Subsequently, the temporally redistributed rewards are further decomposed across agents based on their individual contributions, ensuring that the relationship expressed in~\eqref{eq:rew_red_assumption} is maintained.

Given the dual nature of credit assignment—attributing relative credit to: (1) each time step in the multi-agent trajectory, and (2) each agent at every time step—we can derive the relationship between the trajectory return, $r_{\text{global}, \text{episodic}}$, and the redistributed reward received by agent $i$ at time step $t$, $r_{i,t}$.

To formalize, we assume the following process:

\begin{assumption}
The trajectory return $r_{\text{global}, \text{episodic}}$ is redistributed temporally to obtain rewards for each time step, $r_{\text{global}, t}$, such that:
\begin{equation}
r_{\text{global}, \text{episodic}}(\tau) = \sum_{t=1}^{|\tau|} r_{\text{global}, t}(h_t, a_t, h_{|\tau|}, a_{|\tau|}).
\label{eq: temporal_global_rew_red_assumption}
\end{equation}
Subsequently, these temporally redistributed rewards are decomposed across agents:
\begin{equation}
r_{\text{global}, t}(h_t, a_t) = \sum_{i=1}^N r_{i, t} (h_{i, t}, a_{i, t}, h_{\neg i, t}, a_{\neg i, t}, h_{|\tau|}, a_{|\tau|}),
\label{eq: agent_rew_red_assumption}
\end{equation}
ensuring that:
\begin{equation}
r_{\text{global}, \text{episodic}}(\tau) = \sum_{i=1}^N \sum_{t=1}^{|\tau|} r_{i, t}(h_{i, t}, a_{i, t}, h_{\neg i, t}, a_{\neg i, t}, h_{|\tau|}, a_{|\tau|}).
\label{eq: temporal_agent_rew_red_assumption}
\end{equation}
\end{assumption}

This two-step decomposition process—first temporally and then across agents—ensures that each agent's contribution at each time step is appropriately credited, thereby providing a robust framework for reward redistribution in multi-agent systems.


Let's define a function $w_t \sim W_{\omega}(h_t, a_t, h_{|\tau|}, a_{|\tau|}): (\mathcal{H} \times \mathcal{A}) \times (\mathcal{H}_{|\tau|} \times \mathcal{A}_{|\tau|}) \to \mathbb{R}$ that redistributes the rewards across the temporal axis of the multi-agent trajectory. 
Thus, we can express the multi-agent temporal reward at an arbitrary time step $t$ as
\begin{equation}
r_{\text{global}, t} = w_t r_{\text{global}, \text{episodic}}(\tau)
\end{equation}

Similarly, let's define a function $w'_{i,t} \sim W_{\kappa}(h_{i,t}, a_{i,t}, h_{\neg i, t}, a_{\neg i, t}): \mathcal{H}_i \times \mathcal{A}_i \times \mathcal{H}_{\neg i} \times \mathcal{A}_{\neg i} \to \mathbb{R}$ that redistributes the temporal rewards at an arbitrary time step $t$ across agents. 
Hence, now we can express the reward that agent $i$ receives as
\begin{equation*}
r_{i, t} = w'_{i,t} r_{global, t}
\end{equation*}

Finally, deriving the relationship between $r_{i,t}$ and $r_{\text{global}, \text{episodic}}(\tau)$ for an arbitrary time-step $t$
\begin{equation}
r_{i, t} = w'_{t, i} w_t r_{\text{global}, \text{episodic}}(\tau)
\end{equation}


Based on the definition of reward redistribution function

\begin{align}
    \sum_{i=1}^N \sum_{t=1}^{|\tau|} r_{i, t}(h_{i,t}, a_{i,t}, h_{\neg i, t}, a_{\neg i, t}, h_{|\tau|}, a_{|\tau|}) = r_{\text{global}, \text{episodic}}(\tau) \nonumber \\
    (\sum_{i=1}^N \sum_{t=1}^{|\tau|} w'_{t, i} w_t r_{\text{global}, \text{episodic}}(\tau) = r_{\text{global}, \text{episodic}}(\tau) \nonumber \\
    \sum_{i=1}^N \sum_{t=1}^{|\tau|} w'_{t, i} w_t = 1 \nonumber \\
    \sum_{t=1}^{|\tau|} (\sum_{i=1}^N w'_{t, i}) \times w_t = 1 \nonumber \\
\end{align}

The solution for the above equation is,
\begin{equation}
    \label{eq:agent_weights}
    \sum_{i=1}^N w'_{t, i} = 1
\end{equation}
\begin{equation}
    \label{eq:temporal_weights}
    \sum_{t=1}^{|\tau|} w_t = 1
\end{equation}


To construct the new reward function $\mathcal{R}_{\omega, \kappa}$, we incorporate the original reward $\mathcal{R}_{\zeta}$ and the redistributed credit $r_{i,t}$ that each agent $i$ receives at time step $t$. This redistributed credit reflects the agent's relevance to the final outcome of the multi-agent system. The relevance of each state-action tuple is determined by the reward redistribution functions $W_{\omega}$ and $W_{\kappa}$, which effectively assign the trajectory rewards temporally and across agents.

\begin{align}
    \label{eq:new_rew_func_def}
    R_{\omega, \kappa}(s_t, a_t, s_{t+1}) = R_{\zeta}(s_t, a_t, s_{t+1}) + r_{i, t} \nonumber \\
    R_{\omega, \kappa}(s_t, a_t, s_{t+1}) = R_{\zeta}(s_t, a_t, s_{t+1}) + w_{t, i} w_t r_{\text{global}, \text{episodic}}(\tau)
\end{align}

This formalizes the reward redistribution process, ensuring that each agent receives a reward proportionate to its contribution to the overall team performance at each timestep.

\subsection{Optimal Policy Preservation}
\label{subsec:optimal_policy_preservation}
To ensure that the optimal policy learned using the densified reward function is also optimal in the environment's original reward function, we establish the following theorem:

\begin{theorem}
    Let's consider two Dec-POMDPs as defined in subsection~\ref{subsec:decpomdp_def}, $\mathcal{M}_{\text{env}} = (\mathcal{S, A, P, T, O, N}, \mathcal{R}_{\zeta}, \rho_0, \gamma)$ and $\mathcal{M}_{\text{rrf}} = (\mathcal{S, A, P, T, O, N}, \mathcal{R}_{\omega, \kappa}, \rho_0, \gamma)$. The only distinction between $\mathcal{M}_{\text{env}}$ and $\mathcal{M}_{\text{rrf}}$ are the reward functions. If $\pi_{\theta}^*$ is the optimal policy in $\mathcal{M}_{\text{rrf}}$ then $\pi_{\theta}^*$ is also optimal in $\mathcal{M}_{\text{env}}$.
\end{theorem}

\begin{proof}
    We know that $\pi_{theta}^*$ is optimal in $\mathcal{M}_{\text{rrf}}$. For $\pi_{theta}^*$ to be optimal in $\mathcal{M}_{\text{env}}$, we need to show that $\mathcal{R}_{\omega, \kappa} = \mathcal{R}_{\zeta} + \mathcal{F}(s_t, a_t, s_{t+1})$ where $\mathcal{F}(s_t, a_t, s_{t+1})$ is a potential based shaping function which is a necessary and sufficient condition for optimal policy preservation \ref{subsec:potential_based_reward_shaping}. 

    It is therefore sufficient to show that the equation~\eqref{eq:new_rew_func_def} takes the form $\mathcal{R}_{\omega, \kappa}(s_t, a_t, s_{t+1}) = \mathcal{R}_{\zeta}(s_t, a_t, s_{t+1}) + \gamma \phi(s_{t+1}) - \phi(s_t)$. Comparing this format to equation~\eqref{eq:new_rew_func_def}, assuming $\gamma=1$ we arrive at $\phi(s_{t+1}) - \phi(s_t) = w'_{ t, i} w_t r_{\text{global}, \text{episodic}}(\tau)$. This relation holds for $\phi(s_t) = r_{\text{global}, \text{episodic}}(\tau) (\sum_{t'=0}^{t} w'_{t', i} w_t')$
\end{proof}

This result ensures that if an arbitrary policy $\pi_{\theta}$ when trained using the reward function $\mathcal{R}_{\omega, \kappa}$ in $\mathcal{M}_{\text{rrf}}$ converges to an optimal policy $\pi_{\theta}^*$ then $\pi_{\theta}^*$ will also be optimal for the original reward function $\mathcal{R}_{\zeta}$ in $\mathcal{M}_{\text{env}}$.

\subsection{Policy Gradient Update Equivalence with Reward Redistribution}
\label{subsec:pg_update_eq}

In this subsection, we establish that the policy gradient update for an arbitrary agent $k$, derived from the reward redistribution function, shares the same direction but exhibits a smaller magnitude than the policy gradient update in the environment's original reward function. Furthermore, this ensures that the policy update trajectory towards the optimal policy for an arbitrary initial policy is preserved for every agent.

\begin{proposition}
Let $\pi_{\theta}$ be the parametric policy in a decentralized execution paradigm, where the joint policy is expressed as a product of individual agent policies $\pi_{\theta} = \prod_{k=1}^N \pi_{\theta_{k}}$ \citep{Oliehoek2016ACI, amato2024partial}. The policy gradient update for an arbitrary agent $k$ under the reward redistribution function $R_{\omega, \kappa}$ follows the same direction as the policy gradient update under the environment's original reward function $R_{\zeta}$, preserving the policy update trajectory towards the individual optimal policies.
\end{proposition}

\begin{proof}
Consider the policy gradient update for agent $k$ under the reward redistribution function:
\begin{align}
\nabla_{\theta_{k}} \mathbb{E}_{\pi_{\theta_{k}}}\left[\sum_{t=1}^{|\tau|} r_{k, t}\right] = \nabla_{\theta_{k}} \mathbb{E}_{\pi{\theta_{k}}}\left[\delta(\tau) r_{\text{global}, \text{episodic}}(\tau)\right] = \nabla_{\theta_{k}} \mathbb{E}_{\pi{\theta_{k}}}\left[\sum_{i=1}^N \sum_{t=1}^T \delta_{t} r_{i, t}\right],
\end{align}
where $\tau$ is the multi-agent trajectory attained from the joint policy $\pi_{\theta}$ and $\delta : \mathcal{H} \times \mathcal{A} \times \mathcal{H}_{|\tau|} \times \mathcal{A}_{|\tau|} \to \mathbb{R} \ge 0$ is a non-negative scalar-valued function conditioned on the trajectory. For brevity, we drop the detailed notation of the variables.
From the definition of the reward redistribution function in Assumption~\ref{assumption: rew_red_ass}, we have:
\begin{align}
    \nabla_{\theta_{k}} \mathbb{E}_{\pi_{\theta_{k}}}[r_{\text{global}, \text{episodic}}(\tau)] &= \nabla_{\theta_{k}} \mathbb{E}_{\pi_{\theta_{k}}}\left[\sum_{i=1}^N \sum_{t=1}^T r_{i, t}\right] \\
    &= \nabla_{\theta_{k}} \mathbb{E}_{\pi_{\theta_{k}}}\left[\sum_{t=1}^T r_{k, t} + \sum_{i \neq k} \sum_{t=1}^T r_{i, t}\right] \\
    &= \nabla_{\theta_{k}} \mathbb{E}_{\pi_{\theta_{k}}}\left[\sum_{t=1}^T r_{k, t}\right] + \nabla_{\theta_{k}} \mathbb{E}_{\pi_{\theta_{k}}}\left[\sum_{i \neq k} \sum_{t=1}^T r_{i, t}\right].
\end{align}

Given the definitions \( \sum_{i=1}^N w'_{t, i} = 1 \), equation~\eqref{eq:agent_weights}, and \( \sum_{t=1}^{|\tau|} w_t = 1 \), equation~\eqref{eq:temporal_weights}), we rewrite the above equation as:
\begin{align}
    \nabla_{\theta_{k}} \mathbb{E}_{\pi_{\theta_{k}}}[r_{\text{global}, \text{episodic}}(\tau)] &= \nabla_{\theta_{k}} \mathbb{E}_{\pi_{\theta_{k}}}\left[\sum_{t=1}^{|\tau|} r_{k, t}\right] + \nabla_{\theta_{k}} \mathbb{E}_{\pi_{\theta_{k}}}\left[\left(\sum_{t=1}^{|\tau|} w_t (1 - w'_{k, t})\right) r_{\text{global}, \text{episodic}}(\tau)\right] \\
    \text{Let} \; (w_t (1 - w'_{k, t})) &= M_{t}, \\    \nabla_{\theta_{k}} \mathbb{E}_{\pi_{\theta_{k}}}[r_{\text{global}, \text{episodic}}(\tau)] &= \nabla_{\theta_{k}} \mathbb{E}_{\pi_{\theta_{k}}}\left[\sum_{t=1}^{|\tau|} r_{k, t}\right] + \nabla_{\theta_{k}} \mathbb{E}_{\pi_{\theta_{k}}}\left[\left(\sum_{t=1}^{|\tau|} M_{t}\right) r_{\text{global}, \text{episodic}}(\tau)\right] \\
    \nabla_{\theta_{k}} \mathbb{E}_{\pi_{\theta_{k}}}\left[\left(1 - \sum_{t=1}^{|\tau|} M_{t}\right) r_{\text{global}, \text{episodic}}(\tau)\right] &= \nabla_{\theta_{k}} \mathbb{E}_{\pi_{\theta_{k}}}\left[\sum_{t=1}^{|\tau|} r_{k, t}\right] \\
    \text{Comparing with} \; \nabla_{\theta_{k}} \mathbb{E}_{\pi_{\theta_{k}}}\left[\sum_{t=1}^{|\tau|} r_{k, t}\right] &= \nabla_{\theta_{k}} \mathbb{E}_{\pi_{\theta_{k}}}\left[\delta(\tau) r_{\text{global}, \text{episodic}}(\tau)\right], \; \delta(\tau) = 1 - \sum_{t=1}^{|\tau|} M_{t}.
\end{align}

Since \( 1 \geq (1 - w'_{k, t}) \geq 0 \), \( 1 \geq w_t \geq 0 \), and \( \sum_{t=1}^{|\tau|} w_t = 1 \), the term \( \sum_{t=1}^{|\tau|} w_t \times (1 - w'_{k, t}) \) represents a weighted sum, ensuring that:

\begin{align}
    1 \geq \delta(\tau) &= 1 - \sum_{t=1}^{|\tau|} M_{t} \geq 0.
\end{align}

Thus, training a policy with the reward redistribution function is equivalent to training with the environment's original reward function, as it preserves the direction of each agent's policy gradient update. This ensures that the policy evolves similarly in both settings and that the policy update trajectory for an arbitrary initial policy is preserved.
\end{proof}

\subsection{Temporal-Agent Reward Redistribution (TAR$^2$) architectural details}
\label{subsec:TAR$^2$_arch}
We build upon the architecture proposed by \citet{xiao2022agent} by extending its capabilities to decompose the episodic reward (trajectory return) not only temporally but also at the agent level. This advancement allows our model to predict $r_{i,t}$, effectively learning implicit temporal and agent weights. These weights satisfy the relationship $r_{i,t} = w'_{t, i} w_t r_{\text{global}, \text{episodic}}(\tau)$, thereby enhancing the accuracy and granularity of credit assignment in multi-agent systems.

\section{Experimental Setup}
We demonstrate the effectiveness of our approach TAR$^2$ with single-agent and multi-agent reinforcement learning algorithms against some competitive baselines in the 5m\_vs\_6m battle scenario of the SMACLite \citep{michalski2023smaclite} environment.

\subsection{Baselines}
\begin{figure}[htbp]
    \centering
    \begin{subfigure}[b]{0.45\textwidth}
        \centering
        \includegraphics[width=\textwidth]{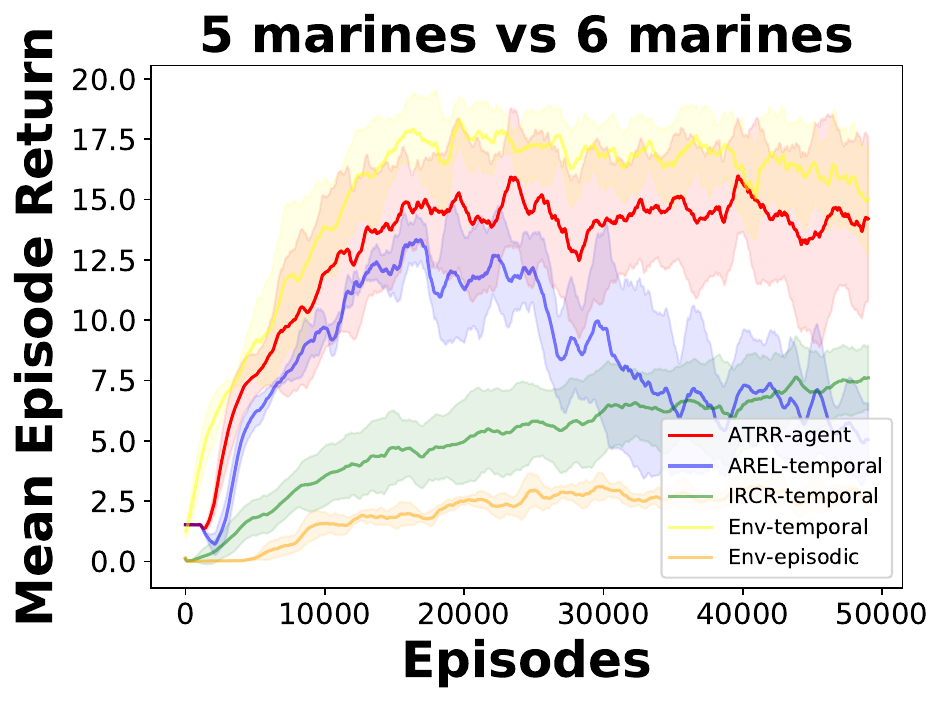}
        \caption{Performance of IPPO in different reward settings. }
        \label{subfig:IPPO_5m_vs_6m}
    \end{subfigure}
    \hfill
    \begin{subfigure}[b]{0.45\textwidth}
        \centering
        \includegraphics[width=\textwidth]{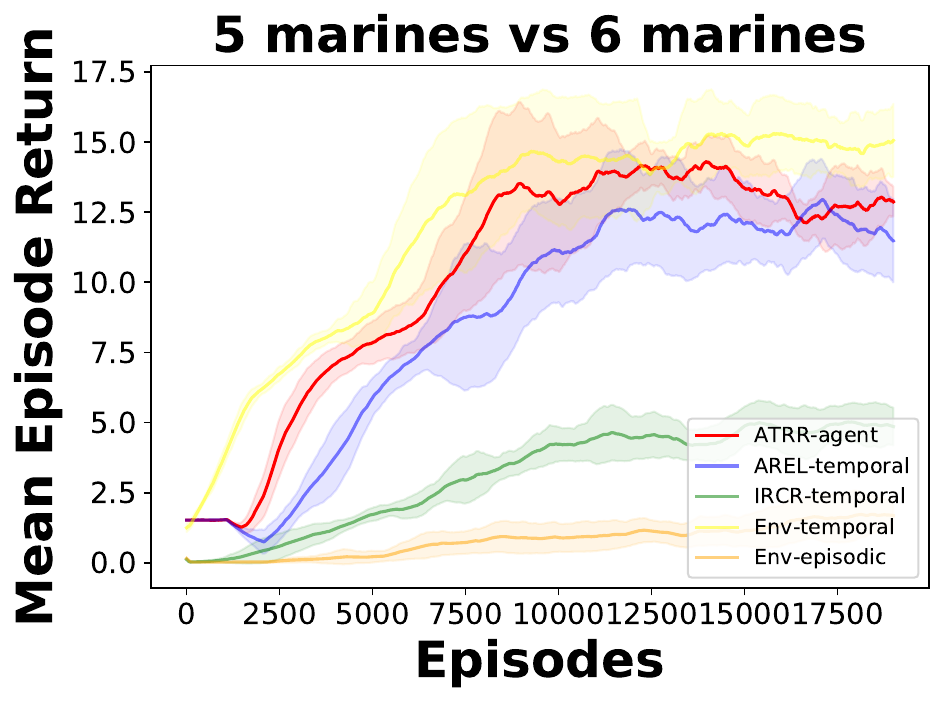}
        \caption{Performance of MAPPO in different reward settings.}
        \label{subfig:MAPPO_5m_vs_6m}
    \end{subfigure}
    \caption{Average agent episodic rewards with standard deviation for task 5m\_vs\_6m.}
    \label{fig:Performance_5m_vs_6m}
\end{figure}

In order to validate the effectiveness of our reward redistribution mechanism we compare its performance with many other forms of reward functions. We train all the baseline reward functions with IPPO\citep{de2020independent, schulman2017proximal} and MAPPO \citep{yu2022surprising} and report them in Fig~\ref{fig:Performance_5m_vs_6m} 

\textbf{Episodic rewards}: This is the episodic reward setting where each agent receives a global reward signal at the end of the trajectory.

\textbf{Dense temporal rewards}: In this setting, each agent receives the original global dense reward signal described in subsection~\ref{subsec:env}.

\textbf{Dense AREL temporal rewards}: This setting employs AREL reward redistribution that temporally assigns rewards to the multi-agent trajectory as described in \citep{xiao2022agent}.

\textbf{Dense IRCR temporal rewards}: In this setting, each agent receives a global reward at every time step following this equation $r_{global, t} = R_{episodic}(\tau)/|\tau|$  \citep{gangwani2020learning}. This baseline also exemplifies a unique form of reward redistribution in our case where the state and action tuple of each agent is of equal importance and hence each of them receive the same global reward.

\textbf{Dense TAR$^2$ agent rewards (ours)}: This is the reward setting proposed in the subsection~\ref{subsec:TAR$^2$_arch}.

\subsection{Environment}
\label{subsec:env}
\textbf{StarCraft Multi-Agent Challenge Lite (SMAClite)}: 
The StarCraft Multi-Agent Challenge (SMAC) \citep{samvelyan2019starcraft} is an RL environment based on the StarCraft II real-time strategy game, in which a team of agents fights against an opposing team controlled by the game engine's centralized hard-coded AI.  We specifically consider the the lightweight and open-source SMACLite version \citep{michalski2023smaclite}.  We consider a battle scenario, 5m\_vs\_6m, where 5 agent-controlled marines battle 6 enemy marines. In this battle situation, the dense reward received by a particular agent while attacking an enemy unit is the difference in the health and shield points removed from that enemy unit in that particular timestep. If a particular agent kills an enemy unit, it receives a reward of 10. Upon defeating the entire enemy team, a reward of (200 / number of agents) is given to each surviving agent. The returns are then normalized such that the maximum possible group return is 20. However, we acummulate the dense reward for each multi-agent trajectory and provide it as a feedback only at the end of the episode.

\section{Results and Discussion}
As presented in Figure~\ref{fig:Performance_5m_vs_6m}, our method TAR$^2$-agent outperform other reward function baselines except for the environment's original dense reward setting as described in subsection~\ref{subsec:env}. This particular baseline is an oracle since it has been manually designed to achieve the objective of this specific environment. While training TAR$^2$, we used the same hyperparameters as proposed in \citep{xiao2022agent} with a slight modification to the training procedure. Since AREL \citep{xiao2022agent} was trained with off-policy reinforcement learning algorithms like QMIX \citep{rashid2020monotonic}, they seemed to not require a warm-up period to train the reward function alone. Since in our experiments we train single and multi-agent on-policy policy gradient algorithms, we empirically discovered that a warm-up period (2000 episodes) performed better.

\section{Conclusion and future work}
This paper studied the multi-agent agent-temporal credit assignment problem in MARL tasks with episodic rewards. We proposed a agent-temporal reward redistribution (TAR$^2$) function that theoretically guarantees the preservation of the optimal policy under the original reward function. Our experimental results demonstrate that TAR$^2$ outperforms all baselines, showing faster convergence speed.

In future work, we want to explore the agent-temporal reward redistribution by utilizing the attention weights generated by the temporal and agent attention blocks during a forward pass since they naturally fit well in the proposed framework. We want to also demonstrate the effectiveness of our approach against more competitive state-of-the-art baselines and across a variety of other MARL environments of varying difficulty. An interesting line of investigation would be to see the transfer-learning capabilities of such models 1) with more agents than it was trained with 2) across different environments with similar objectives.



\bibliography{main}
\bibliographystyle{rlc}






\end{document}